# Graphene-Assisted Chemical Stabilization of Liquid Metal Nano Droplets for Liquid Metal Based Energy Storage

Afsaneh L.Sanati [1], Timur Nikitin [2], Rui Fausto [2,3], Carmel Majidi [4], Mahmoud Tavakoli [1*]

[1] Institute of Systems and Robotics, Department of Electrical and Computer Engineering University of Coimbra, Coimbra 3030–290, Portugal

[2] Centro de Química de Coimbra-Institute of Molecular Sciences (IMS), Dept. de Química, Universidade de Coimbra, Portugal

[3] Faculty of Sciences and Letters, Dept. of Physics, Istanbul Kultur University, Ataköy Campus, Bakırköy 34156, Istanbul, Turkey

[4] Soft Machines Lab, Department of Mechanical Engineering, Carnegie Mellon University, Pittsburgh, PA 15213, USA

[*] Corresponding Author, Email: Mahmoud@isr.uc.pt (Mahmoud Tavakoli)

**Abstract**

Energy storage devices with liquid metal electrodes have attracted interest in recent years due to their potential for mechanical resilience, self-healing, dendrite-free operation, and fast reaction kinetics. Gallium alloys such as Eutectic Gallium Indium (EGaIn) are particularly interesting due to their low melting point and high theoretical specific capacity. However, when exposed to a highly alkaline electrolyte, thin-film electrodes composed of EGaIn droplets are not stable due to the dissolution of Gallium oxide. In this letter, we address this bottleneck by introducing chemically stable films in which nanoscale droplets of EGaIn are coated with trace amounts of graphene oxide (GO). We demonstrated that a GO to EGaIn weight ratio as low as 0.01 provides enough protection for a thin film formed by GO@EGaIn nanocomposite against significantly acidic or alkaline environments (pH 1-14). We show that the GO coating significantly enhances the surface stability in such environments, and thus improves the energy storage capacity by over 10 times. We perform detailed microstructural analysis of various GO@EGaIn composites and analyze their electrochemical cycling performance under various conditions. We then use this technique to fabricate a thin-film supercapacitor. The results indicate that when coating the EGaIn by a GO to EGaIn ratio of 0.001, the areal capacitance is improved by 10x, reaching 20.02 mF/cm$^2$. This discovery can lay the foundation for the development of the next generation of high-performance liquid metal based thin-film electrodes for energy storage applications.

**Introduction**:

Gallium-based liquid metals (LMs) such as eutectic indium gallium (EGaIn) have emerged as a popular material in soft-matter engineering due to their unique combination of high electrical conductivity ($\sigma = 3.4\times10^4$ Scm$^{-1}$) and fluidic deformability, resilience against mechanical strain, and their self-healing properties[1-2]. These LMs are finding their way into a wide range of applications, including soft-matter electronics [3-5], health monitoring patches [6-7], Robotics [8-9], and printed antennas [10]. More recently, they have received increasing attention as energy storage materials [11-13]. This is due to the inherent benefits in the fluidic deformability of LMs, including dendrite-free operation, self-healing properties, high mechanical resilience, and fast reaction kinetics [14]. In contrast to more conventional liquid metal batteries [15-16] that must be heated to operate with liquid phase electrodes, room temperature LMs (RTLMs) like EGaIn are especially attractive because they maintain their liquid phase at room temperature and can be used in devices that are inherently soft and mechanically deformable. Moreover, gallium has a high theoretical capacitance, which makes it an attractive candidate for high power density applications in wearable computing and soft robotics. The following formula can be used to get the theoretical capacitance:

$C = n \times F/M \times V$.

where n is the number of electron transfers, F is the Faraday constant, M is molecular mass, and V is the applied potential. For the case of Ga, n = 3, and M is 69.723 gr/mol. Therefore, the theoretical capacity is 1153.2 mAh g$^{-1}$.

Moreover, Gallium based liquid metals can be easily recycled from the composites, as shown in recent works[13] . Additionally, other liquid metal alloys such as GaIn, GaSn, and GaZn, [17-18], can be studied for improving the areal capacitance and long-term stability. These properties make liquid metal electrodes very attractive options for energy storage, which are being explored by many research groups and industries. However, until now, a key limitation has been their reactivity with alkaline electrolytes, which interferes with their ability to scale down LM to form nanocomposites with high energy density.

Although EGaIn is promising as an electrode material for soft batteries and energy-storage devices, progress in mass scale manufacturing and adoption depends on more robust material architectures and fabrication techniques. This includes the need for metal-electrolyte interactions that can occur at a stable interface over large areas. This is especially challenging with Ga-based

LM alloys since they depend on the formation of a thin surface oxide in order to form stable micro/nanodroplets that can support large LM-electrolyte interfacial areas. Without this oxide skin, the droplets will coalesce and form a single large droplet with poor mechanical stability and a low surface-to-volume ratio. However, the presence of alkaline or acidic electrolytes will cause the LM droplets to lose their surface oxide, resulting in droplet coalescence and a corresponding degradation in the LM-electrolyte interface area and energy storage capacitance.

In fact, in other application contexts, several groups studied the development of printable conductive inks and composites through the transformation of the bulk LM into micro/nano droplets through ultrasonic force [19-22].

However, these microdroplets cannot be directly employed as electrodes in energy storage applications. This is because these droplets are naturally stabilized by an ultrathin (~3nm) [23-24] gallium oxide shell that spontaneously develops on the surface [25]. This protective oxide layer of the Gallium droplets rapidly corrodes in the highly acidic/alkaline electrolytes, thus resulting in the coalescence of the droplets into bigger droplets. Addressing this challenge is key to the development of high-capacity liquid metal electrodes.

In this work, we overcome this key barrier in LM-based energy storage by utilizing LM nanodroplets coated with graphene oxide (GO) shells as electrodes for supercapacitors (SCs). In particular, we discovered that coating EGaIn with trace amount of GO leads to improved stability in electrolytic solutions over the complete range of acidity (pH 1-14). This enables fabrication and application of stable thin-film liquid metals as high capacity electrodes in energy storage devices. In addition to examining the underlying properties of GO-coated EGaIn electrodes, we show for the first-time the ability to create thin-film LM-GO SCs with high areal capacitance. This work is competitive when compared to the few recent liquid metal-based SCs that have used bulk EGaIn.

For instance, in one work [26], a flexible supercapacitor was fabricated by encaging the LM bulk with $Fe_3O_4$ in polyurethane (PU)@polymethacrylate (PMA) fibers. Authors showed that the areal specific capacitance of 3.24 mF/cm$^2$ with scan rate of 200 mV/s. However, in practice it is desirable to produce flat and stable electrodes through techniques such as screen printing that are widely used in fabrication of batteries and Supercapacitors. Moreover, the use of bulk metal in energy storage devices is not optimal due to the low surface area, when compared to micro or nano particles.

By encapsulating the EGaIn microdroplets with graphene oxide (GO), we demonstrate that they can significantly improve their morphological stability in the presence of highly acidic/ alkaline electrolyte. This results in fabrication of electrodes that have high surface to volume ratio, and are stable in aggressive electrolyte environment, which in turn results in a significant improvement of the energy storage capacity. Recent works have shown that negatively charged GO sheets can self-assemble on positively charged EGaIn droplets due to electrostatic charges [27]. When mixed in an acidic environment, galvanic replacement [28-30] occurs as well. In an acidic environment EGaIn droplets interact with hydrogen ions ($H^+$) to produce gallium ions ($Ga^{3+}$). Building on these previous efforts, we create a GO-coated EGaIn (GO@EGaIn) electrode architecture (with a weight ratio of 0.001 GO) that exhibits a 20.02 mF/cm$^2$ areal storage capacity, which is ~10 times higher than bulk EGaIn and 4.5 times higher than GO-only electrodes at a current 0.6mA/cm$^2$. The chemical stability enabled by GO is a function of the GO concentration and the pH of the electrolyte. We prepared and studied 5 composites with different GO-to-EGaIn ratios and studied the reaction of thin-film electrodes made by these composites, when exposed to electrolytes with pH 1-14, through optical and electron microscopy. This permitted the determination of the threshold of GO concentration that enables enough chemical stability in highly acidic/alkaline electrolytes. We further studied the storage capacitance of each of the composites through in-depth electrochemical analysis. As a result, we obtained guidelines for the preparation of GO@EGaIn composites that are suitable for room temperature liquid metal energy storage electrodes.

**Results and Discussion**:

A colloidal solution of GO@EGaIn was prepared by ultrasound-assisted sonication of EGaIn and a water-based GO solution in ethanol (Figure **1A-i**). Five nanocomposites with GO/EGaIn ratio of 0, 0.0001, 0.001, 0.005 and 0.01 were prepared and applied over the glass substrate through spray coating (Figure **1A-ii,**). The resultant thin film electrode was then exposed to HCl and KOH solutions to cover the full range of pH (from 1 to 14). An optical microscope was used to examine how the GO@EGaIn thin films react with electrolytic solutions that have varying pH levels (Figure **1A-iii**). In the case of low surface stability, it is expected that the EGaIn droplets coalesce into larger droplets, thus leaving significant voids on the thin-film (Figure **1A-iii-1**). In the case of stable droplets, the thin-film electrodes are expected to remain intact, as shown in

Figure **1A-iii-2**. In this way, improved chemical stability of nanodroplets within a highly acidic / alkaline solution enables their use as electrodes in energy storage devices.

After finding the GO concentration needed to ensure the morphological stability of the droplets, we developed a symmetric GO@EGaIn supercapacitor (Figure **1B-i**) that exhibits significantly higher energy storage (~14×) compared to the bulk EGaIn (Figure **1B-ii**). Figure **1B-iii** shows a thin-film supercapacitor including four layers. This device was developed by thin-film deposition of GO@EGaIn electrodes. Below the electrode we placed two stretchable current collectors (CC). Specifically, a highly conductive and stretchable ink composed of EGaIn microdroplets, Ag flakes, and styrene-isoprene-styrene (SIS) co-polymer elastomer was used as the main CC and a corrosion-inhibiting composite of carbon black and SIS (CB-SIS) was used as a second CC. The electrode geometry was custom-made through a simple and accessible 1064nm wavelength inferred (IR) laser patterning method [31]. Thanks to the symmetric architecture, the full film of the CCs, and electrodes can be pre-coated and then rapidly turned into the desired shape through laser patterning. Figure **1C** shows a thin-film SC created as an example and depicts how it can be bent, twisted, or rolled. The laser processing technique described here serves as a universal tool for quick, scalable, and inexpensive production of thin-film GO@EGaIn SCs because of its versatility in quickly producing 2D patterns.

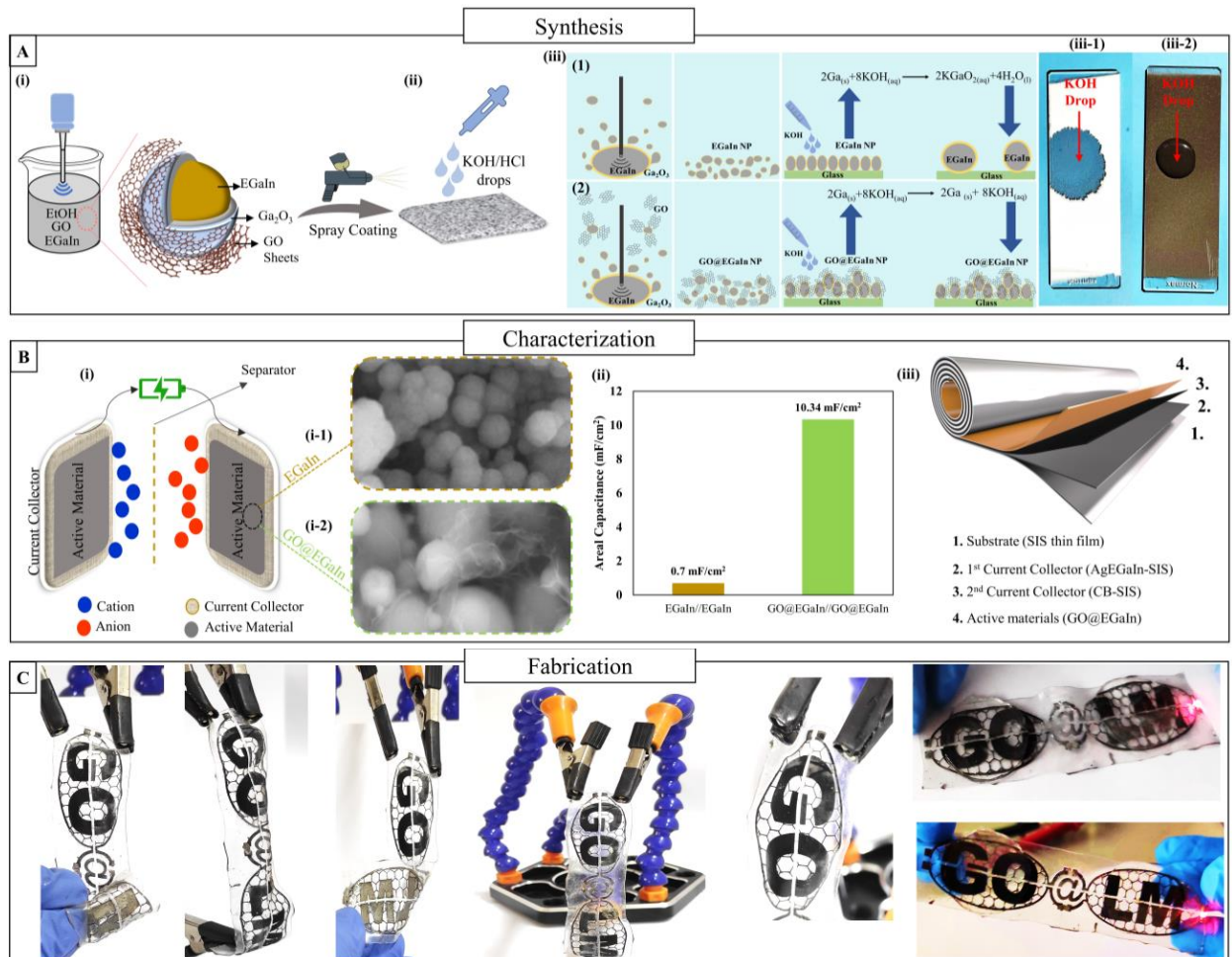

Figure 1. A) Synthesis of GO@EGaIn nanocomposite. (i), Spray coating of GO@EGaIn thin film (ii). The reaction of thin film making by EGaIn (iii-1) and GO@EGaIn (iii-2) in exposure to KOH/HCl solution. B) The schematic of the symmetric supercapacitor used in this study (i), The SEM image of EGaIn electrode (i-1) and GO@EGaIn electrode (i-2), The areal specific-capacitance results obtained from EGaIn and GO@EGaIn electrodes (ii), and the schematic of different layers in supercapacitor (iii). C) Photographs of a GO@EGaIn-based supercapacitor deformed into different shapes.

In Figure 2, optical microscope images show EGaIn subjected to pH values of 1, 13, and 14 either without GO or with different amounts of GO over a glass substrate. All these images were taken with backlighting in order to inspect the light transmission from the film. By observing the formation of micro holes in the thin-film, it was possible to identify instability of the film due to coalescence of EGaIn droplets. The same samples imaged with a dark background are shown in Supplementary **Figure S2**. Additionally, videos **S1**, **S2**, and **S3** show how EGaIn particles with various GO ratios interact with each other and create bulk particles, especially when exposed to very low or high pH levels. As seen in the videos, GO@EGaIn nanocomposite reacts with an alkaline solution (pH = 13 and 14 in videos **S2** and **S3**) significantly more quickly than it does with an acidic solution (video **S1**).

Referring to Figure **2A** and Video **S1**, voids appear in the GO-free EGaIn film within seconds when subjected to HCl with a pH = 1. This is evidence of instability of the EGaIn droplets due to the elimination of Gallium oxide layer, resulting in droplet coalescence and degradation of the thin-film electrode. The same sample exhibits even faster degradation when subject to alkaline solutions of KOH with pHs 13 and 14 (Figures **2B** and **2C,** and videos **S2,** and **S3**). Referring to Figures 2D, E and F, coating the droplets with a GO to EGaIn ratio of 0.0001 by weight is not enough to maintain stability. When subjected to highly acidic/alkaline solutions, the thin-film is rapidly destroyed and the droplets coalesce. One interesting phenomenon that we observed was that the size of the aggregated mass of EGaIn for a sample with 0.0001 wt GO is larger than with the GO-free sample (Figure **2A**, **B**, **C,** and video **S1).** This is because there is not enough GO on the EGaIn surface to completely protect the oxide shell from dissolution. However, the small remaining GO sheets seem to act as a center for the aggregation of liquid metal droplets around them, which is especially visible in Figure **2D**. Referring to Figures **2G**, **H**, and **I**, increasing the graphene oxide content to 0.001 by weight improves the stability of the nanoparticles and prevents aggregation or coalescence in solvents with pHs of 1, 13, and 14. This is quite visible when comparing the optical images with samples of 0 and 0.0001 wt GO. Nevertheless, even in the case of 0.001 wt GO, some light is still observed to pass through the thin layer, suggesting some aggregation of LM droplets.

Further increasing the GO to 0.005, results in a denser network. The thin film created by this nanocomposite can be observed in Figures **2J** and **K** to be resistant to pHs 1 and 13. This seems to be the right threshold that prevents the dissolution of GO from offering enough protection for the composite in very acidic conditions at pH=13, the same thing occurs. However, in a very alkaline solution with a pH of 14, a 0.005 wt GO coating seems to be still insufficient to offer full protection from aggregation, as seen in Figure **2L**. Further increasing the GO to 0.01, offers full protection in all cases, as can be seen in Figures **2M, N**, and **O**. Recordings of the thin film response to exposure to solvents with pH values of 1, 13, and 14 are displayed in videos **S1**, **S2**, and **S3**, respectively.

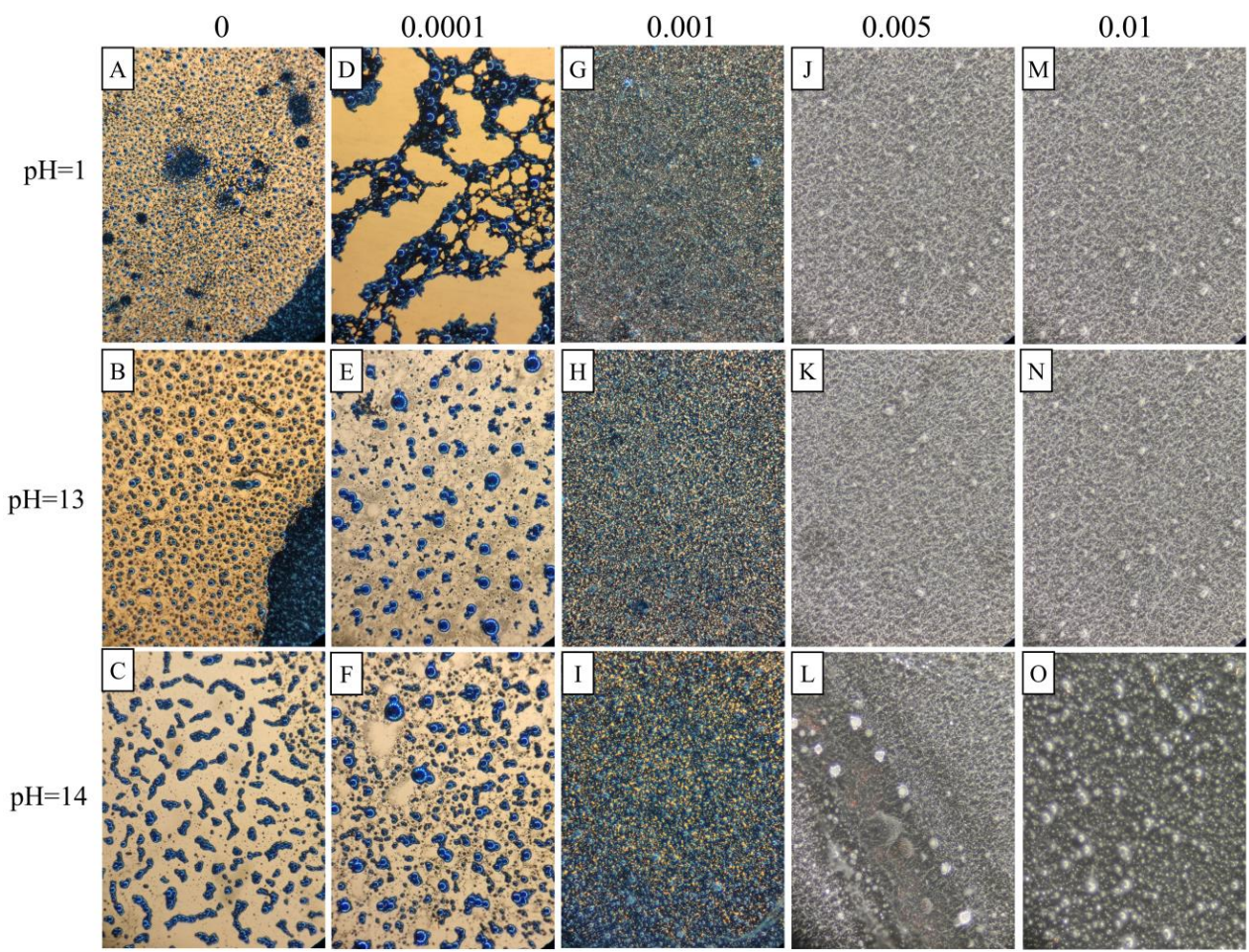

Figure 2. Optical images were taken using a microscope, for GO concentration of 0, 0.0001, 0.001, 0.005 and 0.01 for pHs 1 (A, D, G, J and M), 13 (B, E, H, K and N) and 14 (C, F, I, L and O).

To understand why trace amount of GO makes such large effect on the properties of the composite, we performed an analysis to compare the specific surface of GO and EGaIn droplets. Regarding GO, the specific surface area of GO is known from the literature to be ~2391 m$^2$/g [32]. This is the most similar GO to the one we used (being sonicated to form mono layers and being in an aqueous solution). For LM droplets, we calculated the specific area by first measuring the average size of particles from a SEM image (**S3-A**). Using image J software and drawing a histogram related to particle size, the average particle size was calculated to be around 493.17 nm (**S3-B**). Then, taking into account the spherical shape of the particles, and formulations of the sphere volume and surface, and considering the EGaIn, the specific area of EGaIn droplets was calculated as 1.95 m2/g (**S3-C**). That is, the specific area of GO is approximately 1000x higher than EGaIn nanodroplets. As a result, when GO/EGaIn ratio is only 0.001, the GO surface area equals the surface area of the EGaIn particles. However, this is not yet enough to offer full protection to the EGaIn droplets, as naturally not all GO sheets coat the LM droplets. However, increasing this by 5x, at a GO@EGaIn weight ratio of 0.005, a significant protection of EGaIN droplets to acidic/alkaline environment is established.

**Table 1** shows a summary of the reaction of the thin-film electrodes with different GO/EGaIn weight ratios for exposure to solvents with pHs ranging from 1 to 14. In the table, the check mark (✓) indicates that the thin films remain intact after exposure to KOH/HCl for enough time until the solution is dried. A cross mark (×) indicates that the thin-film did not remain intact when exposed to solvent.

Table 1: Thin films behaviors with varying amounts of GO and pH

| pH \ GO% | 1 | 2 | 3 | 4 | 5 | 6 | 7 | 8 | 9 | 10 | 11 | 12 | 13 | 14 |
|---|---|---|---|---|---|---|---|---|---|---|---|---|---|---|
| 0 | ✗ | ✓ | ✓ | ✓ | ✓ | ✓ | ✓ | ✓ | ✓ | ✓ | ✓ | ✓ | ✗ | ✗ |
| 0.0001 | ✗ | ✓ | ✓ | ✓ | ✓ | ✓ | ✓ | ✓ | ✓ | ✓ | ✓ | ✓ | ✗ | ✗ |
| 0.001 | ✓ | ✓ | ✓ | ✓ | ✓ | ✓ | ✓ | ✓ | ✓ | ✓ | ✓ | ✓ | ✗ | ✗ |
| 0.005 | ✓ | ✓ | ✓ | ✓ | ✓ | ✓ | ✓ | ✓ | ✓ | ✓ | ✓ | ✓ | ✓ | ✗ |
| 0.01 | ✓ | ✓ | ✓ | ✓ | ✓ | ✓ | ✓ | ✓ | ✓ | ✓ | ✓ | ✓ | ✓ | ✓ |

Overall, when the EGaIn-GO-free thin film is exposed to an alkaline/acidic solution, the oxide passive layer on the surface of the liquid metal ($Ga_2O_3$) will be removed because of the spontaneous redox reaction. The products are very soluble.

In alkaline solution $\qquad Ga_2O_{3\,(s)} + 2KOH_{(l)} \longrightarrow 2K(Ga(OH)_4)_{(l)} + 3H_2O_{(l)}$

In acidic solution $\qquad Ga_2O_{3\,(s)} + 6HCl_{(l)} \longrightarrow 2GaCl_{3\,(l)} + 3H_2O_{(l)}$

Once the oxide layer is removed, the EGaIn droplets join each other and aggregate into large particles, thus reducing the surface area.

However, in the presence of GO, the protons and electrons from an acidic or basic solution first react with GO functional groups. Therefore, if there is sufficient GO to completely cover the EGaIn drops, the passive oxide layer of the LM is protected from dissolution, and the thin film will be intact. In this instance, the GO layers will serve as a barrier between the EGaIn drop and the electrolyte.

Scanning Electron Microscopy (SEM) and Energy-dispersive X-ray spectroscopy (EDS) were performed in order to analyze the surface topology and elemental composition of the thin film nanocomposite, respectively. An analysis was made on two sets of samples. The first set contained 3 samples and were synthesized by sonication of 1 g of LM in ethanol without any

GO, for 20 minutes, 1, and 2 hours, and spray coated over a glass substrate. Four distinct samples were included in the second set, each created with a different amount of GO (0.0001, 0.001, 0.005, and 0.01). The first set was investigated to Figure out the effect of the sonication time on the production of graphitic carbon oxide layers.

Figure **3** shows the film made with GO-free LM droplets, sonicated for 20 minutes (3A, B), 1 hour (3C, D), and 2 hours (3E, F). Increased sonication time reduces the size of the droplets. Interestingly, it can be seen in the higher magnification images (3B, D, F) that even without the addition of GO, a small amount of carbon layer is formed around the particles. The pyrolysis of ethanol and subsequent self-assembly results in the formation of graphite oxide layers. This phenomenon was also reported in [33]. However, it seems that unlike the samples with GO, here this layer is not well distributed everywhere and is concentrated on specific zones. Figure **S4** displays additional SEM images and EDS analysis. Increasing the sonication time to 2-hours led to an increase in the formation of graphite matrix. Furthermore, the higher sonication duration also enhanced the dispersion of LM nanodroplets throughout the graphite matrix (Figure 3F). This can be observed in the SEM images, where a higher density of nanodroplets is clearly visible in the treated zone (Figure 3E). Figure **S6** depicts SEM and backscattering images, as well as the EDS analysis, and Figure **S7** shows color maps for the gallium, indium, and carbon elements.

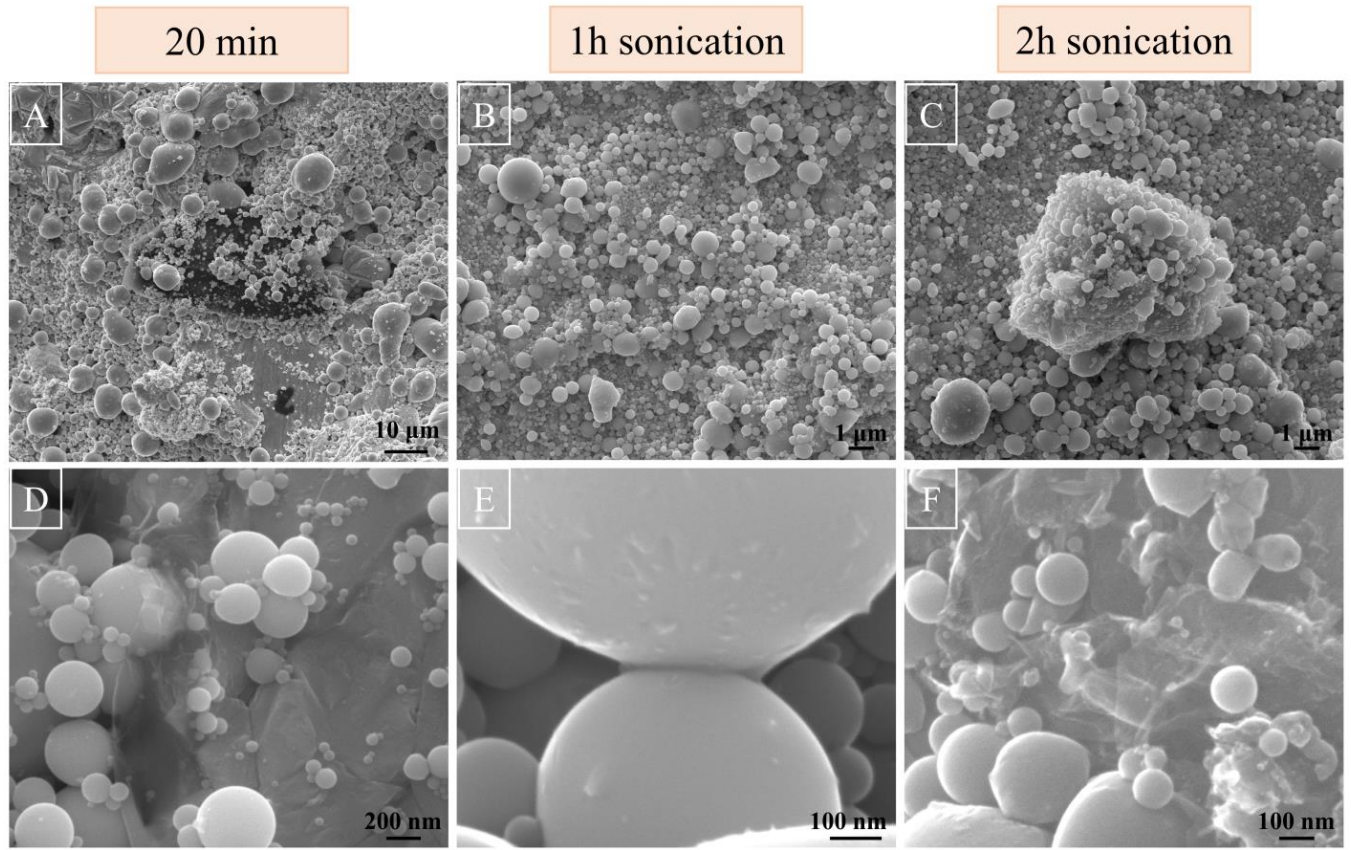

Figure 3. SEM images of LM@EtOH composite synthesis using ultrasonic method with A) 20 minutes, B) 1 hour, and C) 2 hours sonication time. D, E and F) respectively are the magnified of A, B and C images.

Figure **4** shows the SEM images of the second batch, i.e., the GO@EGaIn nanocomposite with varying amounts of GO. In order to synthesize all of these nanocomposites, all of these solutions were subjected to 20 minutes of sonication to avoid excessive creation of graphite layers, as these can prevent proper interaction between LM and GO.

As can be seen in Figures **4A** and **B**, for samples with 0.0001 and 0.001 of GO to EGaIn ratio, some GO sheets are observable, but these didn't fully cover the liquid metal droplets (Figure **4E** and **F** show high magnifications of images **4A** and **B**, respectively). When the GO percentage is increased to 0.005 and 0.01, respectively, the GO sheets are visible all around the particles, resulting in better protection of the LM droplets, (Figures **4C**, **D**, **G** and **H**). It also seems that some of the GO sheets maintain the LM droplets connected to each other. Figure **S8-S12** displays additional SEM and EDS analyses for samples with 0.0001, 0.001, 0.005, and 0.01 percentages of GO respectively.

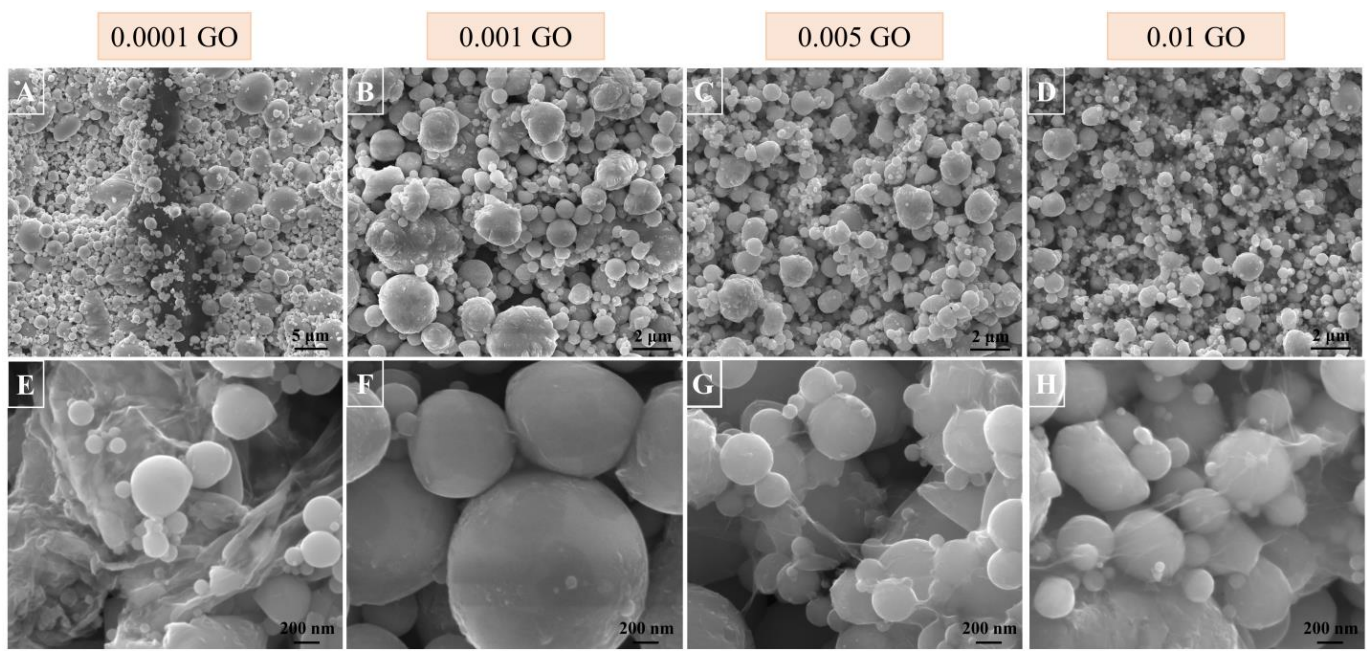

Figure 4. SEM images of GO@EGaIn nanocomposite synthesis with GO to EGaIn ratio of A) 0.0001, B) 0.001, C) 0.005 and D) 0.01 of GO, during the period of 20 minutes of sonication. Images E, F, G and H) respectively are magnified of A, B, C and D photos.

Raman spectroscopy was applied to study carbon sheets prepared by sonication of liquid metal (EGaIn) particles in ethanol with and without the addition of GO. Previous reports showed that under an ultra-sonic cavitation field, ethanol is converted to oxidized graphitic carbon layers on the surface of EGaIn liquid metal particles [33]. In this case, it is possible to underestand, why samples labeled as "EGaIn- GO free" still exhibit resistance to aggregation, and offer some chemical stability. The presence of graphite is manifested by observation of the typical Raman features for this material: prominent bands at ~1345 cm$^{-1}$ (D band; $A_{1g}$) and ~1600 cm$^{-1}$ (G band; $E_{2g}$) [34-35]. According to expectations, sonication of ethanol for 20, 60, and 120 min (samples 1, 2, and 3, respectively) led to the formation of graphene oxide, as shown by Raman spectroscopy, with sample 1 showing the weakest D and G bands (Figure **5**). For samples 5–8, with added GO, the intensity of the D and G bands increases with an increasing amount of added GO (Figure **5-A**).

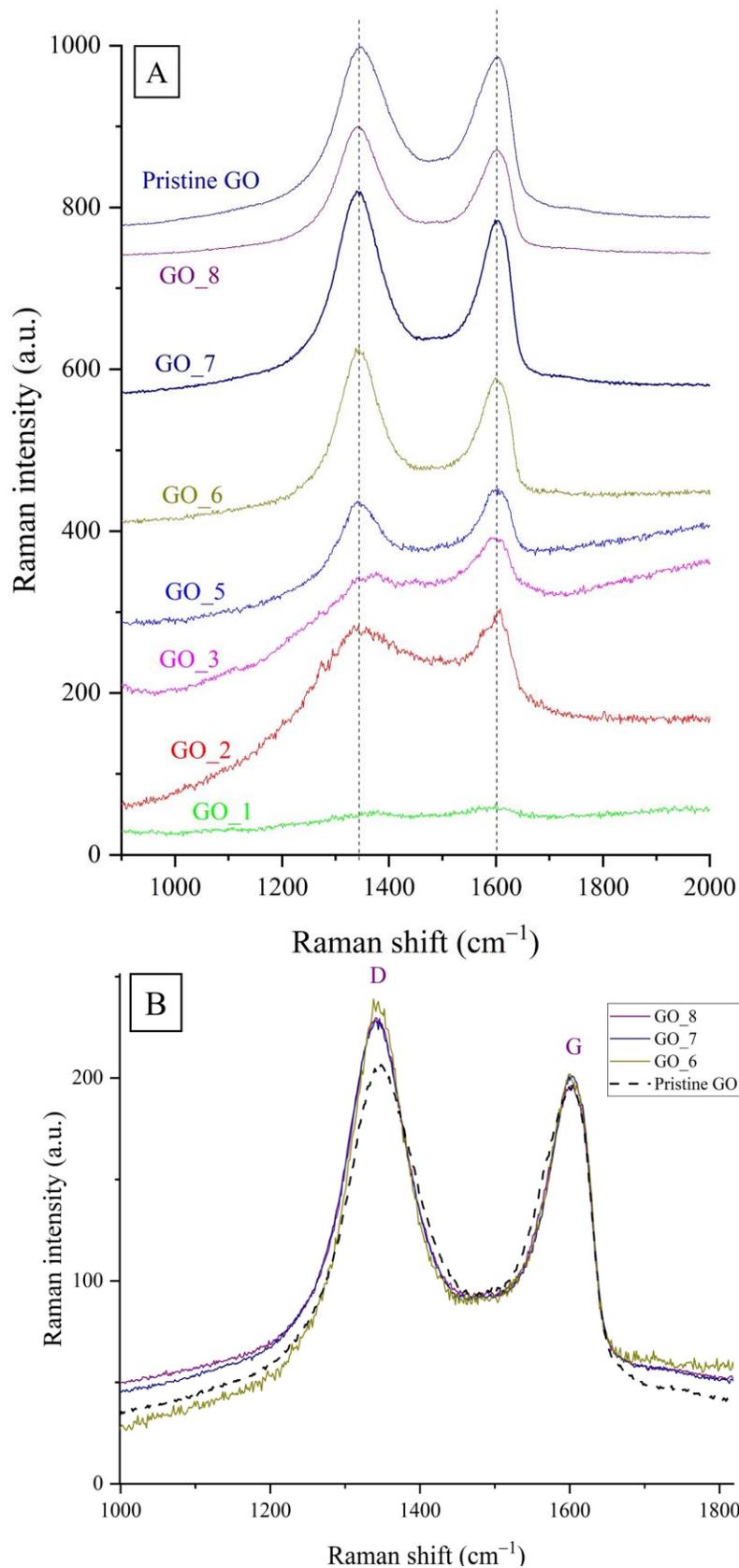

Figure 5. (A) Typical Raman spectra of samples as a function of sonication time (20 – 120 min; GO_1 – GO_3) and amount of added GO (0.0001 – 0.01; GO_5 – GO_8) and (B) Raman spectra of pristine GO and samples 6, 7, and 8 with GO content 0.001, 0.005, and 0.01 respectively, normalized by the G band.

After sonication, the D/G band intensity ratio increases from 1.06, for pristine GO, to 1.26, 1.16, and 1.22, for GO-6, GO-7, and GO-8, respectively (Figure **5-B**). An increase in the D/G intensity ratio is associated with the GO reduction by liquid metals [36], and it may indicate either a decrease in the average size of the $sp^2$ domains upon reduction of GO [37] or the presence of remnant unpaired defects after the partial removal of the oxygen-containing functional groups [38]. The GO reduction by liquid metals for the studied samples, however, seems to not be as effective as that made by other methods like laser annealing [39]. Indeed, it has been reported that the position of the G band is shifted towards lower wavenumbers with the decrease of the oxidation degree [37, 40], while in the case of the studied samples, the position of this band (~1602 cm$^{-1}$) is unchanged.

Electrochemical analysis was performed in order to analyze the electrochemical behavior of the GO@EGaIn in the presence of aggressive electrolytes. Therefore, as part of our study, we prepared four electrode sets with different GO amounts and analyzed their storage capacity by cyclic voltammetry (CV) and galvanostatic charge-discharge (GCD). This includes a GO-free liquid metal electrode, a GO-only electrode as references, and two GO@EGaIn electrodes with GO/LM weight ratios of 0, 0.0001, and 0.001.

The electrochemical performance of these electrodes is summarized in Figure **6**. Each row of the Figure shows the behavior of one of the composites, and the last row shows the comparison between them. The first column, from left to right, displays the cyclic voltammogram of each electrode at various scan rates to determine the electrodes' potential windows. The GCD curves for each electrode with various current densities in the potential window identified by the CV technique are displayed in the second column. The final column displays the areal capacitance at various current densities as determined by GCD curves. In these electrochemical cells, we used 6M KOH as the electrolyte. This is because K$^+$ serves as the cation (and has the second-highest ionic conductivity after H$_3$O), while OH$^-$ is the anion (and has the highest ionic conductivity and mobility in water). Additionally, the small size of the OH$^-$ anion´s improves intercalation, and thus enhances the pseudocapacitive action. Figure **6A** shows the cyclic voltammetry (CV) plots of the symmetric system made using the GO-free EGaIn electrodes at various scan rates (50, 100, 150, 200, 250, 300, 350, 400, 450, and 500 mV/s) between 0 to 0.35 V. The symmetric LM electrode system showed oxidation and reduction peaks at 0.220 and 0.170 V respectively. By increasing the scan rate potential, the oxidation peaks shift to 0.320 V but the reduction peak doesn't shift. According to Pourbaix's diagram, the electrochemical redox process involves the following reactions:

$$Ga^{2+} + 3H_2O \leftrightarrow GaO_3^{3-} + 6H^+ + e^-$$

$$E_a = 1.868 - 0.354pH + 0.0591 \log \frac{[GaO_3^{3-}]}{[Ga^{2-}]}$$

Figure **6B** displays the Galvanostatic Charge Discharge (GCD) for the electrode created by an EGaIn droplet at various current densities (0.6, 0.8, 1.0, 1.2, 1.4, and 1.6 mA/cm²) in the 0-0.3V range. According to earlier research, typical current densities are between 0.01 and 1 mA/cm². Therefore, we chose the range of 0.6 to 1.6 mA/cm² for characterization of the supercapacitor. The GCD curves in Figure **6B** for all cases exhibit a nonlinear plateau structure, indicating pseudocapacitive electrochemical behavior. As can be seen from curves **6B** and **6C**, the device's charge capacitance increases as the current density decreases, which is determined using the following equation:

$$C = \frac{I\Delta t}{\Delta V}$$

where $I$ is the discharge current density, $\Delta V$ is the potential window, and $\Delta t$ is the discharge time. Due to insufficient time for electrolyte ion diffusion, as illustrated in **Figure 6B**, the discharge time decreases with increasing current density.

**Figures 6D, E,** and **F** show the electrochemical behavior of the electrodes, when adding trace amounts of GO (GO /LM=0.0001). Compared to the GO-free electrodes, the shape of the CV graphs of the GO@EGaIn nanocomposite is observed to be different (Fig. 6D). This should be due to the insertion of an additional reaction on the electrode's surface, which results in shifting the oxidation peaks to 0.470 V and the reduction peak to 0.14 V. The areal capacitances for different current densities are shown in **Figures 6E** and **F**, respectively. Only a slight improvement over the GO-free sample was observed. Similarly, Figure **6G, E, and F**, show the electrochemical behavior for the sample with a GO to LM weight ratio of 0.001. As can be seen, the addition of the GO coating to the LM droplets improves the areal capacitance by approximately 10 times, compared to the GO-free samples. This demonstrates the significant effect of the GO coating on the storage capacity of LM droplets.

The CV curves in Figure **6G** show that GO added to LM droplets significantly changes their shape, taking on a semi-rectangular shape with oxidation and reduction peaks at 0.47 and 0.33V, respectively. The significant improvement of the GO@EGaIn electrode compared to the GO-free sample might originate from the contribution of the pseudocapacitive effect or electrical double-layer capacitors (EDLC) to the overall specific capacity. To verify this, we have calculated the individual contribution of the EDLC mechanism to the overall specific capacity by studying the electrochemical behavior of a GO-only electrode. **Figure 6J** displays CV graphs that show the performance of GO//GO systems in the potential range of 0 to 0.85V with various scan rates. As shown in the graphs, these systems exhibit an ideal non-faradic behavior. Here, the GO was reduced using a fiber laser as detailed in the Experimental Section, in order to make the electrodes less electrically resistive. However, because of the presence of the GO@EGaIn composite, the GO is partially reduced by the gallium oxide replacement and sonication power [36, 41]. As expected, the LM-free electrode demonstrates a pure EDLC behavior. This is also apparent from the symmetry of the GCD curves when plotted for different current densities (Figure **6K**) within a voltage window of 0–0.85 V. This confirms a reversible ion adsorption/desorption process at the surface of GO electrode. As shown in Figure **6L**, the amount of specific capacitance decreased from 4.44 to 4.04 mF/cm$^2$ by increasing the current densities from 0.6 to 1.6 mA/cm$^2$. We note that here the amount of GO is ~70× higher than what exists in GO@EGaIn electrode with a GO/LM weight ratio of 0.001. Therefore, the amount of GO in the GO@EGaIn electrodes is too low to contribute toward increasing the capacitance of the electrode, due to the EDLC behavior. Therefore, the increased capacitance seems to be purely related to the improved morphological stability of the EGaIn microdroplets.

Figure **6M** illustrates the comparison between the CV curves of all electrodes at a 100 mV/s scan rate of potential. Figure **6N** shows the discharge time of all electrodes, and Figure **6O** shows the areal specific capacitance of the EGaIn// EGaIn of electrodes based on the GCD curves at a current density of 1 mA/cm$^2$. The GO@EGaIn electrode shows a significant improvement compared to both the GO-free LM electrode and EGaIn-free GO electrode. Compared to EGaIn without GO, the areal capacitance is increased by ~15×. This clearly shows the effect of the improved surface area of the LM droplets and the surface stability offered by the encapsulation through GO nanosheets. In fact, the areal capacitance of GO can also contribute to the areal capacitance of GO@EGaIn. However, it is important to note that only trace amounts of GO are present in the composite, and thus the improvement is mostly related to the improved surface area of EGaIn droplets. As the amount of graphene oxide continues to increase, as shown in Figure **S13-A, B**, the areal capacitance increases. Also, the zoomed image of the GCD graph is also displayed, which indicates that the potential drop is around 0.3V in all cases (Fig. **S13-A**).

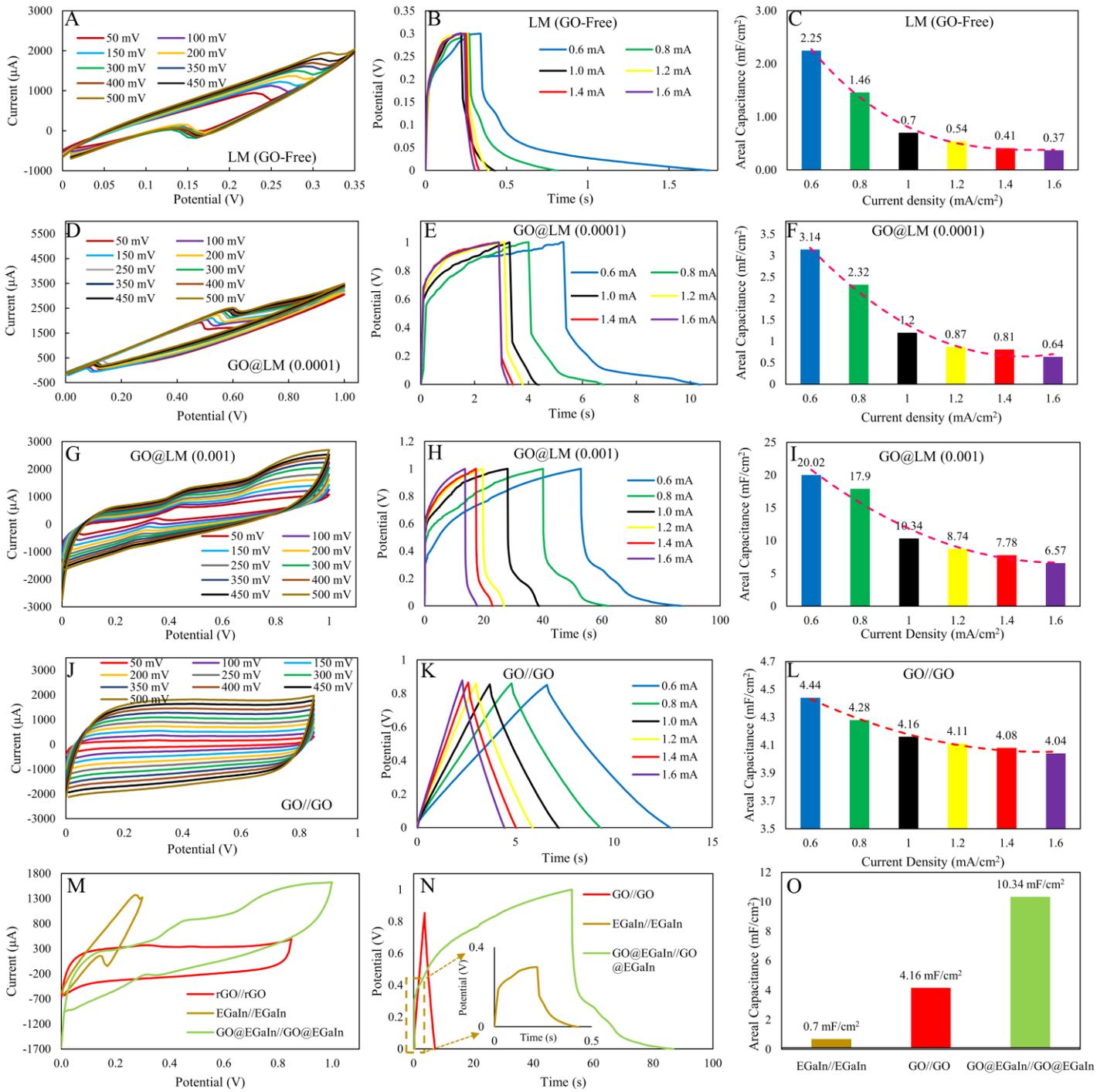

Figure 6: A, B and C) CV and GCD curves of EGaIn//EGaIn and its areal capacitance versus current densities, respectively. D, E and F) CV and GCD curves of GO@EGaIn with 0.0001 of GO and its areal capacitance versus current densities, respectively. G, H and I) CV and GCD curves of GO@EGaIn with 0.001 of GO and its areal capacitance versus current densities, respectively. J,K and L) CV and GCD curves of rGO//rGO and its areal capacitance versus current densities, respectively. M, N and O) Comparison between CV, GCD and the areal capacitance of symmetric EGaIn, rGO and GO@EGaIn electrodes at 100mV and 1 mA/cm$^2$, respectively.

**Application of LMNPs in energy storage:**

We developed a thin-film energy storage device using thin-film deposition and laser patterning to demonstrate an application of the GO@EGaIn nanocomposite (**Figure 7**). This device is composed of two current collector layers: Ag-EGaIn-SIS and CB-SIS. First, a 300µm-thick PVA sacrificial layer is deposited on glass using a thin-film applicator **(7A-ii)**. Next, the PVA layer is spray-coated with a thin GO@EGaIn electrode **(Fig. 7A-iii)**. A thin film applicator is then used to deposit a pair of current collector layers composed of CB-SIS (2$^{nd}$ CC) and Ag-EGaIn-SIS (1$^{st}$ CC) **(7A-iv and v)**. The 1$^{st}$ CC layer provides excellent electrical conductivity and mechanical resilience when subjected to mechanical strain, while the 2$^{nd}$ CC provides chemical protection against corrosion in a highly alkaline electrolyte. This four-layer architecture is then patterned using a Master Oscillator Power Amplifier (MOPA) laser with an infrared wavelength (1064 nm) **(7A-vi)**. Afterward, we coated the resulting device with an 800 µm thick SIS layer **(7A-vii)**. This layer infiltrates, binds to the fabricated thin-film electrodes and CCs, and serves as the elastic substrate for the energy storage device. Next, the multi-layered patch is peeled from the glass surface by a spatula **(7A-viii)** and then immersed in water to dissolve the sacrificial PVA layer **(7A-ix)**. **Figure 7B** shows the design of the electrode, and **Figure 7C** shows a snapshot of the laser processing setup. A photograph of the fabricated device with a Light-Emitting Diode (LED) is presented in **Figure 7D**. The system's charging and discharging process is displayed in Video **S4.**

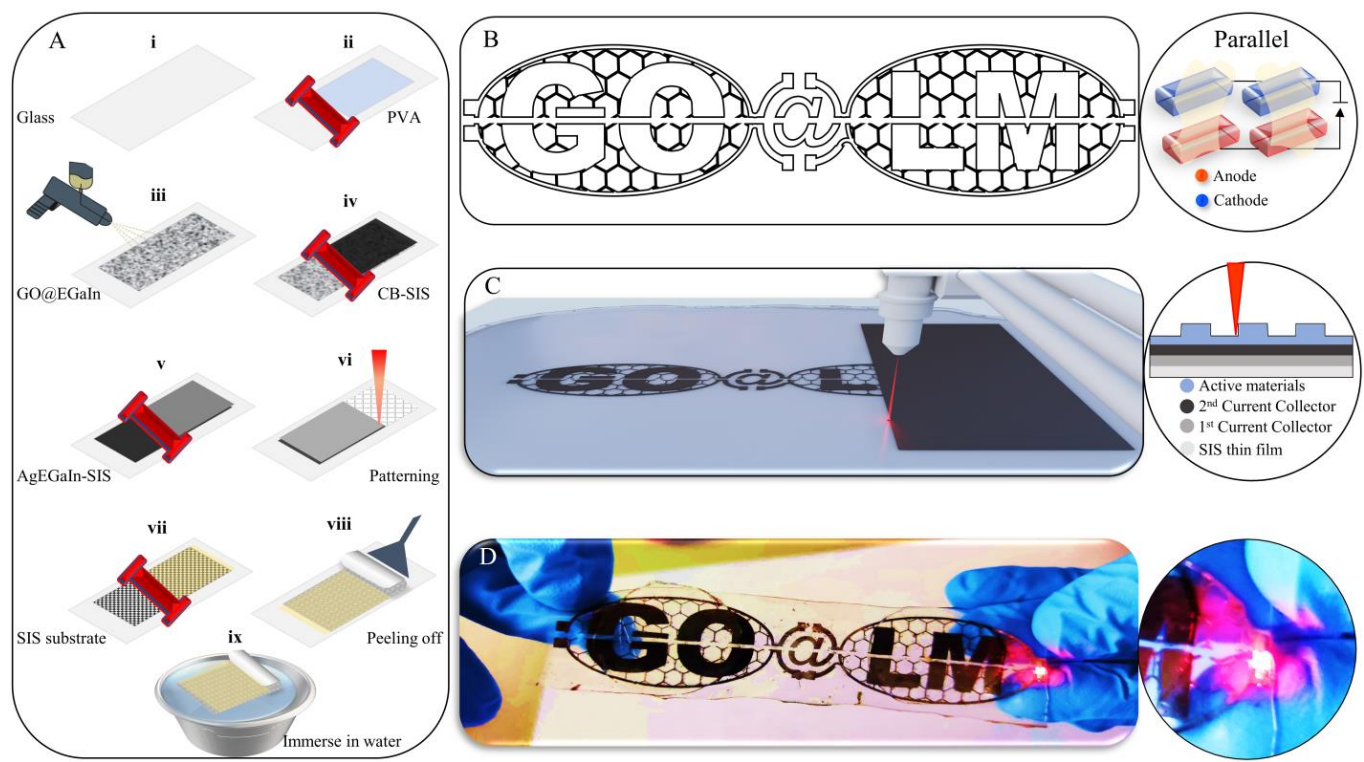

Figure 7: A) Supercapacitor manufacturing steps. B) The first laser-engraving design. C) Schematic of MOPA laser processing. D) Photograph of a symmetric SC powering an LED light.

## Conclusion:

In this study, we demonstrated a technique for maintaining a stable suspension of EGaIn nanodroplets within an electrolytic solution through the use of a graphene oxide coating. We showed that by coating the EGaIn droplets with trace amounts of GO, it is possible to develop thin films that are morphologically stable in strong acidic and alkaline solutions, thus permitting the use of liquid metal thin-film electrodes with a high surface-to-volume ratio for applications in energy storage. When compared with the bulk EGaIn, the GO@EGaIn thin-film electrode can store 10× more energy. We made a detailed study of the GO@EGaIn nanocomposite with different amounts of GO to find the optimal amount of GO that offers enough chemical stability to the GO@EGaIn nanocomposite, when exposed to highly acidic or alkaline solutions. This study was performed through detailed optical and electron microscopy, further supported by Raman spectroscopy. In this study, we showed that GO coating acts as a barrier between EGaIn nanodroplets and protons and electrons from acidic and alkaline electrolytes. As a result, the thin film created by GO@EGaIn even when exposed to extremely acidic or basic solutions, will stay intact.

We found that a GO/EGaIn weight ratio of 0.005 is enough to offer chemical stability to the samples. We further performed an electrochemical study of the thin-film electrodes with different GO amounts and analyzed the CV graphs and their corresponding energy storage capacity. Results show that samples with a GO/EGaIn weight ratio of 0.001 provide significantly higher storage capacity. Investigations demonstrated that symmetric supercapacitors formed with EGaIn//EGaIn electrodes have a capacitance of 2.25 mF/cm$^2$ at currents of 0.6 mA/cm$^2$, which dramatically increases to 20.02 mF/cm$^2$ when adding 0.001 wt of GO. Furthermore, adding this weight fraction of GO results in a 10× and 15× increase in capacitance at 0.6 mA/cm$^2$ and 1.0 mA/cm$^2$, respectively. It causes 15 times more capacitance when compared with a GO-free electrode. This increase in capacitance occurs due to the improved stability of the EGaIn droplet morphology in the presence of KOH, which thus prevents LM particles from joining each other and reduces the surface area of the particles. Finally, we demonstrate the application of this composite for energy storage by creating a thin-film supercapacitor composed of GO@EGaIn electrodes. The supercapacitor is produced using a laser-based patterning method.

The work presented here is the first study to examine the use of graphene-coated liquid metal droplets in energy storage applications. These early results are promising and suggest that GO@EGaIn electrodes could be used as thin-film electrodes that remain intact and maintain a high surface-to-volume ratio in the presence of highly alkaline or acidic solutions. However, further progress depends on a more complete understanding of the influence of particle size and changes to particle composition on morphological stability and charge storage. There are also opportunities to explore the addition of elastomers or gels that can function as an elastic binder, as well as the use of other deposition and patterning methods.

**Experimental Section:**

**Synthesis GO@EGaIn nanocomposite:**

The GO@EGaIn nanocomposite was produced in this study using an ultrasonic technique. As shown in Figure **S14-A**, a total of 5 samples with different percentage mass ratios of GO (0, 0.0001, 0.001, 0.005, and 0.01) were prepared. Herein, ultrasonication time was defined at 20 min as the typical time to produce the GO@EGaIn nanocomposite.

It appears that the final product in the first sample (0 of GO and 1g of LM in 20 ml of ethanol) was created via the mechanism depicted in Figure **S14-B**. In this sample, the solution was subjected to sonication after 1 g of LM was added to ethanol (Fig. **S14-B-i**). Some of the ethanol molecules were pyrolyzed as a result of the sonication irradiation (Fig. **S14-B-ii**), and after further sonication, these molecules joined to form carbon sheets (Fig. **S14-B-iii**), which, as can be seen in Figure **S14-B-iv**, can attach to the oxide layer surrounding the surface of the LM and grow upward due to their functional groups [33]. It seems that in other samples with different amounts of GO, as shown in Figure **S14-C**, another mechanism is conceivable.

In these samples, after mixing LM and GO with ethanol and subjecting the resulting solution to sonication, hydrogen ions first react with the gallium oxide layer ($Ga_2O_3$) on the LM surface due to the presence of acid in the graphene solution (Fig. **S14-C-i**), and after the oxide layer is removed, gallium ions are exposed, and the LM surface becomes positively charged (Fig. **S14-C-ii**). Then, with the assistance of the galvanic replacement mechanism, the layers of GO self-assembled over the LM surface (Fig. **S14-C-vi**) to create the GO@EGaIn nanocomposite because the GO sheets have now become negatively charged as a result of the presence of functional groups in the acid solution due to the electrostatic reaction (Fig. **S14-C-iii**) [28].

**Fabrication of the 2$^{nd}$ current collector (CB-SIS):**

In a 30ml glass flask, 1.1 gr SIS and 15 mL of toluene were added, the flask was then put on the mixer at 500 rpm for 30 minutes. Next, 0.6 grams of CB were added, with the solution being put on the mixer for 3 minutes at 2000 rpm. This mixing cycle allows for the full integration of the conductive material in the polymer mixture, resulting once again in a conductive stretchable ink.

**Fabrication of the 1$^{st}$ current collector (Ag-EGaIn-SIS):**

In a flask, 0.6 g of SIS and 2.4 g of toluene were mixed in the mixer at 500 rpm for 30 minutes. For each gram of solvent solution 1.65 gr, of Ag flakes were added, with the solution being put on the mixer for 3 minutes at 2000 rpm. This was followed by the addition of 1.65 grams of EGaIn for each gram of solution and the posterior mixing of the mixture for 3 minutes at 2000 rpm on the mixer. Both of these mixing cycles allow for the full integration of the conductive materials in the polymer mixture, resulting in a conductive stretchable ink.

**Acknowledgments**

This work is financed by the European Union under a European Research Council grant (ERC, Liquid3D, 101045072). Views and opinions expressed are, however, those of the author(s) only and do not necessarily reflect those of the European Union or the European Research Council. Neither the European Union nor the granting authority can be held responsible for them.

Financing also came from the SMART Display project (reference: POCI-01-0247-FEDER-047153), co-financed by the European Regional Development Fund, through Portugal 2020 (PT2020), and by the Competitiveness and Internationalization Operational Program (COMPETE 2020).